\documentclass[preprint, journal]{vgtc}
\onlineid{0}
\usepackage{xcolor}
\newcommand{\rev}[1]{\textcolor{blue}{#1}}
\usepackage[normalem]{ulem}
\renewcommand{\rev}[1]{#1}
\renewcommand{\sout}[1]{}

\title{\rev{Mitigating Confirmation Bias through Hand-Drawing Videos}}

\author{%
    \authororcid{Chenyu Lin}{0009-0005-7108-7427},
   \authororcid{Cindy Xiong}{0000-0002-1451-4083},
   \authororcid{Icy Zhang}{0000-0003-3423-6794}
}

\authorfooter{
    \item
  	Chenyu Lin is with the University of Wisconsin–Madison.
  	E-mail: cheney.lin@wisc.edu
    \item
    Cindy Xiong is with Georgia Institute of Technology.
  	E-mail: cxiong@gatech.edu
    \item
    Icy Zhang is with the University of Wisconsin–Madison.
    E-mail: icy.zhang@wisc.edu
}

\abstract{Understanding data visualizations is essential for informed decision-making, yet interpretation is often shaped and even distorted by prior beliefs. 
We investigate whether an embodied pedagogical approach, in which viewers observe the dynamic hand-drawing of a visualization, can mitigate confirmation bias and improve interpretation accuracy.
We conducted a study comparing static bar charts to videos in which charts are constructed through hand-drawing, across contexts that either align with or challenge participants’ prior beliefs. 
\rev{The results indicate that hand-drawn videos helped participants accurately interpret data, even when the data conflicted with their prior beliefs. } 
This approach also reduced belief-consistent errors and increased belief-overriding responses.
\rev{These findings suggest that exposing the construction process of a visualization supports more accurate reasoning and mitigates the influence of confirmation bias. Consequently, this work introduces a promising design space for bias-mitigating data interfaces.}
}


\graphicspath{{figs/}{figures/}{pictures/}{images/}{./}} 

\usepackage{tabu}                      
\usepackage{booktabs}                  
\usepackage{lipsum}                    
\usepackage{mwe}                       
\usepackage{ccicons}                   
\usepackage{diagbox}                   
\usepackage{mathptmx}                  
\usepackage{float}

\begin{document}


\firstsection{Introduction}

\maketitle

Confirmation bias remains \rev{a} challenge \rev{to accurate data interpretation because people may favor evidence that aligns with prior beliefs while discounting contradictory information. This is especially problematic in visual data communication, where charts are often assumed to support objective reasoning but can instead reinforce existing beliefs}~\cite{li2025confirmation,lisnic2025visualization}.

In this paper, we approach the challenge of mitigating confirmation bias in visualization through the lens of embodied cognition. Our approach is \rev{also} informed by dual-process theories of reasoning, which distinguish between fast, intuitive judgments (System 1) and slower, more deliberate reasoning (System 2)~\cite{kahneman2011thinking}. 

We hypothesize that \rev{hand-drawing} construction of visualizations can reduce biased interpretation through two mechanisms. 
First, embodied construction influences our intuitive (System 1) processing by grounding interpretation in sensorimotor experience, promoting heuristics that better reflect the underlying data~\cite{fan2023drawing}. 
Second, the step-by-step unfolding of the visualization engages analytical (System 2) processing by slowing down interpretation and making intermediate states of the visualization visible, helping viewers to evaluate the visualized patterns more deliberately rather than relying on initial impressions, which could mitigate bias~\cite{xiong2022reasoning}.

To test our hypothesis, we conduct \rev{an initial} study to evaluate whether observing hand-drawing videos of visualizations improves interpretation accuracy and mitigates confirmation bias. 
We also examine whether these effects are stronger in belief-relevant contexts and whether they are associated with reductions in belief-consistent errors and increases in belief-overriding responses.


\rev{Because this intervention combines both embodied and dynamic elements, it evaluates their joint impact as a holistic paradigm rather than isolating their individual effects. We treat this work as a foundational test of an embodiment-inspired visualization technique that establishes a baseline that enables future research to disentangle the specific contributions of embodiment versus step-by-step animation.}


\vspace{2mm}
\noindent \textbf{Contributions:}
We introduce an embodied cognition approach for visualization as a potential intervention to mitigate confirmation bias.
We also provide empirical evidence that observing the step-by-step construction of charts can improve interpretation accuracy, particularly in belief-relevant contexts.

\section{Related Work}
Well-designed visualizations empower readers to extract information through perceptual inference, actively guiding attention and shaping how data are understood \cite{ware2000information, munzner2014visualization}. 
\rev{However, visualization interpretation involves more than perception alone; it also includes mapping visual features to underlying values and integrating this information with prior knowledge~\cite{carpenter1998model, pinker1990theory, shah2002graph}. Because readers often default to intuitive judgments based on salient visual cues rather than systematically evaluating underlying data relationships~\cite{evans2008dual, stanovich2000individual, pinker1990theory, carpenter1998model, shah2002graph}, visualizations can still lead to misinterpretation and cognitive biases when deeper sense-making is required~\cite{franconeri2021science}.}

\subsection{Confirmation Bias}
Confirmation bias refers to the tendency to seek out and interpret information in ways that align with one's existing beliefs \cite{nickerson1998confirmation}. 
\rev{\sout{This happens because people often rely on heuristics that steer them toward belief-consistent interpretations rather than following the structure of the data~\cite{klein2017sources}.}}
Many factors, such as ambiguity of the task, time pressure, task context, and expertise, can contribute to confirmation bias~\cite{chaiken1994heuristic, kahneman2011thinking}.
\rev{From the perspective of dual-process theories of reasoning ~\cite{chaiken1994heuristic},}
when initial interpretations align with prior beliefs, \rev{individuals} are more likely to accept them quickly through intuitive processing, reinforcing confirmation bias~\cite{nickerson1998confirmation, kunda1990motivated}.
Even when individuals attempt to reason more carefully, early heuristic cues can continue to shape subsequent thinking \cite{chaiken1994heuristic, chang2025early}.
\rev{In visualization contexts, this interaction between prior beliefs and heuristic cues is especially important because visual cues can direct attention to particular patterns, reinforcing viewers' prior beliefs~\cite{li2025confirmation, xiong2022belief}.}


Existing work has explored various approaches to mitigating confirmation bias in data visualization, including improving visual design~\cite{xiong2022reasoning}, incorporating uncertainty representations~\cite{chuanromanee2022crowdsourced}, and providing explanatory~\cite{li2025confirmation} or interactive support~\cite{lisnic2025visualization} to elicit a slower analytical process. 
For example, asking readers to externalize prior expectations in visual form before inspecting data can improve inference quality by enabling direct comparison of the priors with observed evidence \cite{koonchanok2023belief}.
However, these approaches primarily operate at the level of external representation, such as improving visualization design, rather than addressing the cognitive processes that give rise to biased interpretation. 
As a result, even well-designed visualizations may fail to prevent misinterpretation \cite{franconeri2021science}.
This limitation highlights the need for interventions that more directly engage the cognitive processes underlying data interpretation.

\subsection{Embodied Cognition}
Embodied cognition is a view of cognition positing that cognitive processes are grounded in sensorimotor experiences \cite{barsalou2008grounded, wilson2002six}. \rev{Both performing and, importantly, observing physical actions, such as gesture and object manipulation, can enhance understanding by activating shared perceptual–motor representations that simulate the underlying concept, externalizing cognitive processes, and reducing cognitive load \cite{barsalou1999perceptual, barsalou2008grounded, borghi2011embodied, clark2008supersizing, zhang2021instructed, zhang2022watching, zhang2025toward}.} 
\rev{Building on this theoretical framework, visualization researchers began designing systems that leverage direct manipulation and physical actions~\cite{lee2021data}. For example, a study by Zhu et al.~\cite{Zhu2023Cognitive} on data physicalization demonstrates that embodied interaction with data representations can enhance reasoning and understanding.} \rev{Lee et al.~\cite{lee2021data} introduced data visceralization, using virtual reality to help viewers develop a more bodily and experiential understanding of abstract measures in data.} 

\rev{Embodied approaches, such as observing a hand dynamically construct a visualization, may likewise improve visualization interpretation.
Rather than focusing only on the most visually salient elements, such as the tallest bar in a chart~\cite{xiong2022reasoning}}, viewers may attend to how values are progressively constructed and compared, emphasizing relationships, scaling, and proportionality. More broadly, constructing or interacting with representations has been shown to promote deeper cognitive processing by encouraging learners to organize and integrate information~\cite{fiorella2015generative, vanmeter2005drawing, zhang2025drawing}.


Hand-drawing is a practical form of embodied pedagogy in which visual representations are created through physical action~\cite{fan2023drawing}. 
\rev{Both producing drawings \cite{fiorella2018drawing} and observing an instructor draw \cite{fiorella2016eight, fiorella2019instructor, quillin2015drawing} have been shown to improve conceptual understanding. Recent work further suggests that observing hand-drawing can support learning of statistical concepts and their visualizations. For example, Zhang et al.~\cite{zhang2025drawing} found that participants who watched an instructor draw histograms and normal distributions better understood both the visualizations and the underlying concepts than those viewing static or non-embodied dynamic visualizations. However, prior work has primarily treated visualization as a means of learning domain concepts rather than as an object of learning itself. This study takes a step toward bridging embodied cognition and visualization by investigating whether observing hand-drawing videos can directly support visualization interpretation. }
Less is known about whether embodied approaches can support understanding of visualizations themselves as objects of learning. 
This study takes a step toward bridging that gap between visualization research and embodied cognition.

\section{The Study}
\rev{We examine whether dynamically constructed visualizations improve interpretation accuracy compared with static visualizations (H1).}
We also examine whether these effects depend on the context in which the data are presented, particularly in belief-relevant settings where prior beliefs may bias interpretation (H2). 
Finally, \rev{we explore whether dynamic hand-drawing and gesturing shifts participants’ reasoning processes (H3). Specifically, we test whether it reduces \textbf{belief-consistent errors}, in which participants select an incorrect interpretation that aligns with their prior beliefs, and increases \textbf{belief-overriding responses}, in which participants correctly interpret the data despite their prior beliefs suggesting the opposite conclusion.}

\subsection{Participants}
We recruited undergraduates from a public university. 
Ninety-five participants started the Qualtrics online survey\cite{qualtrics}, and 87 participants finished it. 
Participants received extra credit toward their course grade for participation. 
Participants ranged in age from 18 to 45 years ($M = 21.61$, $SD = 5.61$). Among them, 62 identified as female and 24 as male. Regarding race and ethnicity, 67 participants identified as Hispanic (77.0\%), 8 as Asian or Pacific Islander (9.2\%), 4 as Black or African American (4.6\%), 4 as multiracial or other (4.6\%), 3 as White or Caucasian (3.4\%), and 1 chose not to report (1.2\%). The study design and recruitment procedures were approved by the university’s Institutional Review Board (IRB-20-0419).
 
\subsection{Design \& Procedure}

\begin{figure}[htbp]
  \centering 
  \includegraphics[width=\columnwidth, alt={A procedural diagram illustrating the study workflow: pretest, random assignment to control or experimental condition, two main-task scenarios (without and with confirmation bias), two transfer tasks, and a post-survey.}]{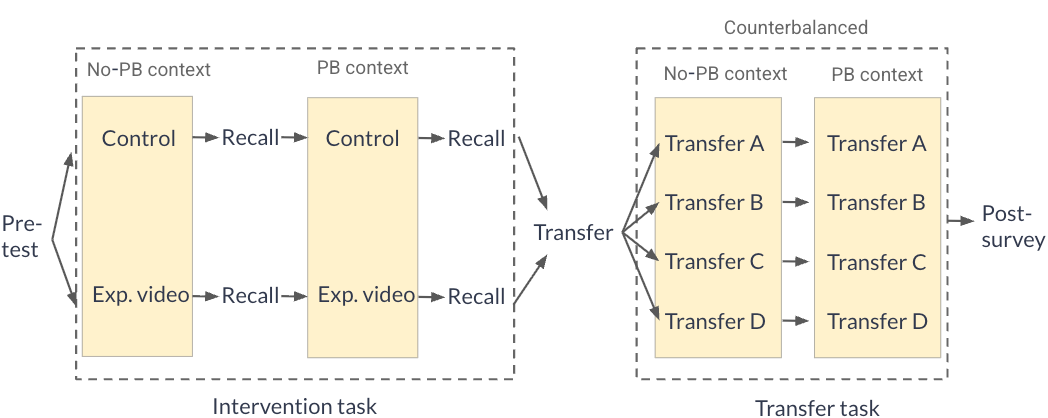}
  \caption{%
  Study design and procedure%
  }
  \label{fig:Study_1_Experiment_Design}
\end{figure}

\begin{figure}[htbp]
  \centering 
  \includegraphics[
    width=0.95\columnwidth,
    alt={A side-by-side comparison of the two conditions: the control condition displaying a static bar graph and the experimental condition showing a dynamically constructed bar graph video.}
  ]{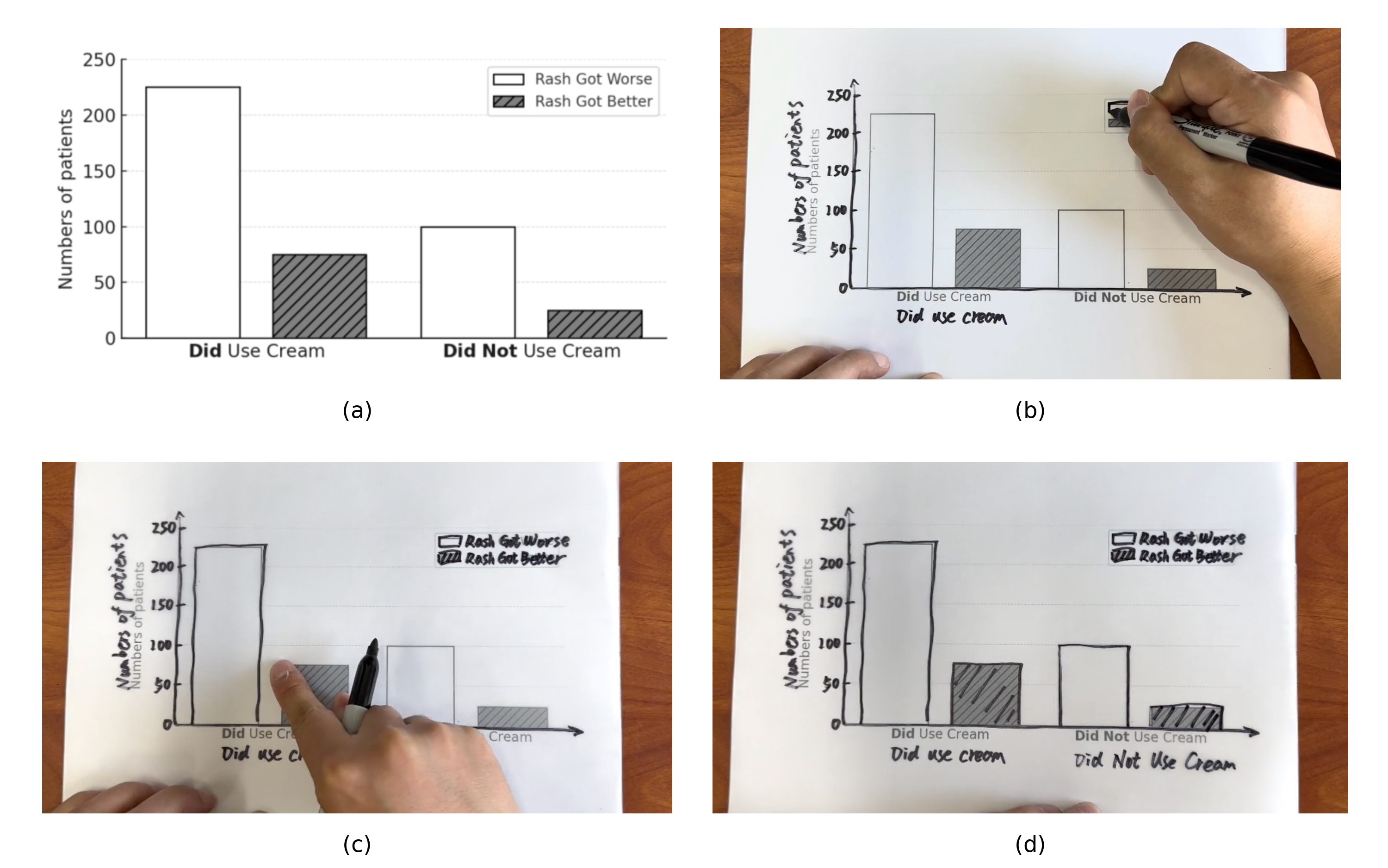}
  \caption{%
  Screenshots of the control (a) and experimental condition (b)--(d)%
  }
  \label{fig:two_conditions}
\end{figure}

The overall experimental workflow is illustrated in \cref{fig:Study_1_Experiment_Design}.
Upon accessing the survey link and providing consent, participants completed a pretest.
They then began the study with an intervention task in which they were randomly assigned to one of two conditions: the control condition ($n = 43$) or an experimental (i.e., embodied) condition ($n = 44$). 
Participants in the control condition read static bar charts, whereas participants in the experimental condition viewed embodied videos of dynamically constructed bar charts (see \cref{fig:two_conditions}).
The embodied video depicted an instructor’s hand constructing a bar graph step by step on tracing paper over a printed template, \rev{without audio. The tracing design was used to ensure precise alignment between the hand-drawn and static visualizations, thereby minimizing unintended differences in graphs across conditions. Only the hand and drawing surface were visible; the instructor's face and body were not shown.} 
The y-axis was introduced first, followed by the x-axis, and then the legend and bars were drawn sequentially. \rev{The videos were played at a fixed accelerated speed so that each video fit within the 90-second exposure window. Playback speed and pacing were held constant across participants in the experimental condition.} 
To prevent participants from skipping the learning materials prematurely and to ensure comparable exposure time across conditions, both intervention pages (video or static image) \rev{required participants to remain for 90 seconds before proceeding.}

\begin{figure}[htbp]
  \centering 
  \includegraphics[width=\columnwidth, alt={Two contextual scenarios. The left one is No-PB context, and the right one is PB context}]{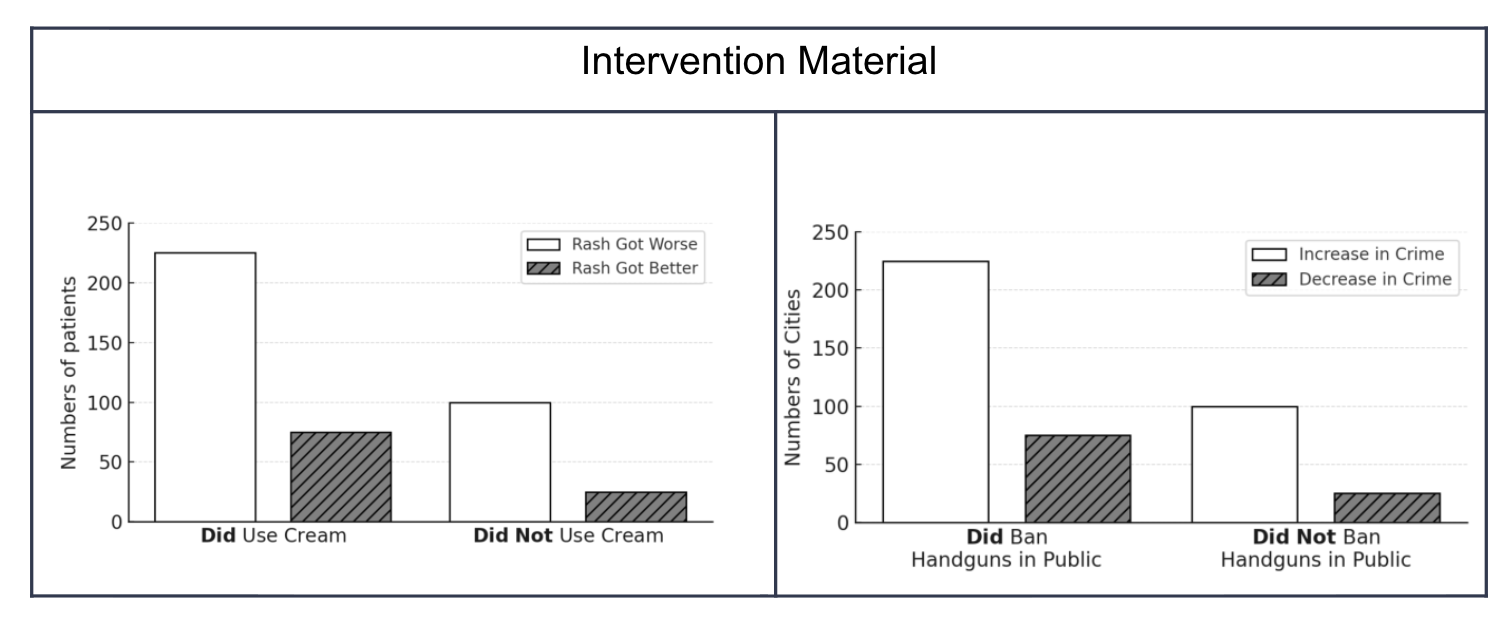}
  \caption{%
  Two contextual scenarios in the intervention task%
  }
  \label{fig:InterventionMateria}
\end{figure}

All participants read two $2 \times 2$  bar charts depicting two scenarios.
The \textbf{first scenario} was designed not to invoke participants’ prior beliefs (PB). 
In this scenario, the bar chart depicted the relationship between the use of a cream and changes in rash severity. We refer to this belief-neutral context as the No-PB (No-Prior Belief) context. 
The \textbf{second scenario} was designed to invoke prior beliefs.
The bar chart depicted the relationship between a policy banning handguns and the crime rate (see \cref{tab:contexts}). \rev{We selected prior-belief topics that were likely to evoke existing beliefs among our participants, recognizing that the salience of these topics may differ across populations.}
We refer to this belief-relevant context as the PB (Prior Belief) context \rev{in the rest of the paper.} 
Although the contextual content differed, the numerical values and visual characteristics of the bar charts (e.g., bar height and color) were kept identical across scenarios. \rev{\sout{but counterbalanced across variables}} See \cref{fig:InterventionMateria}.
\begin{table}[htbp]
\centering
\small
\caption{Context scenarios used in the study}
\label{tab:contexts}

\begin{tabular}{lcc}
\toprule
\diagbox{Context}{Task}& Main & Transfer  \\
\midrule
No-PB & Cream vs. Rash Severity & MOCHAOS Therapy vs. Stress  \\
PB & Gun ban vs. Crime Rates & Vaccination vs. Sickness  \\
\bottomrule
\end{tabular}

\end{table}

After reading the visualization, participants answered a comparison question requiring them to determine the direction of the effect (e.g., `Compared to patients who did not use the cream, patients who used the cream were \textit{more likely} or \textit{less likely} to have their rashes improve').

Next, participants completed a transfer task with two $3 \times 3$  bar charts, first in the No-PB context, followed by the PB context (see \cref{tab:contexts}).
In the No-PB context, the bar graph depicted the relationship between MOCHAOS therapy (a fictional therapy name created for this study) and stress levels. The x-axis indicated whether participants received \textit{MOCHAOS Therapy} or \textit{No MOCHAOS Therapy}, and the y-axis indicated the number of people in each outcome category (\textit{Less Stressed}, \textit{Same}, or \textit{More Stressed}). 
In the PB context, the bar graph depicted the relationship between vaccination and sickness frequency. The x-axis indicated whether participants \textit{Did Vaccinate} or \textit{Did Not Vaccinate}, and the y-axis indicated the number of people in each outcome category (\textit{Getting Sick Less}, \textit{Same}, or \textit{Getting Sick More}). 

\begin{figure}[htbp]
  \centering 
  \includegraphics[width=\columnwidth, alt={Transfer task. Participants were randomly assigned to one of two counterbalanced versions that reversed the numerical values.}]{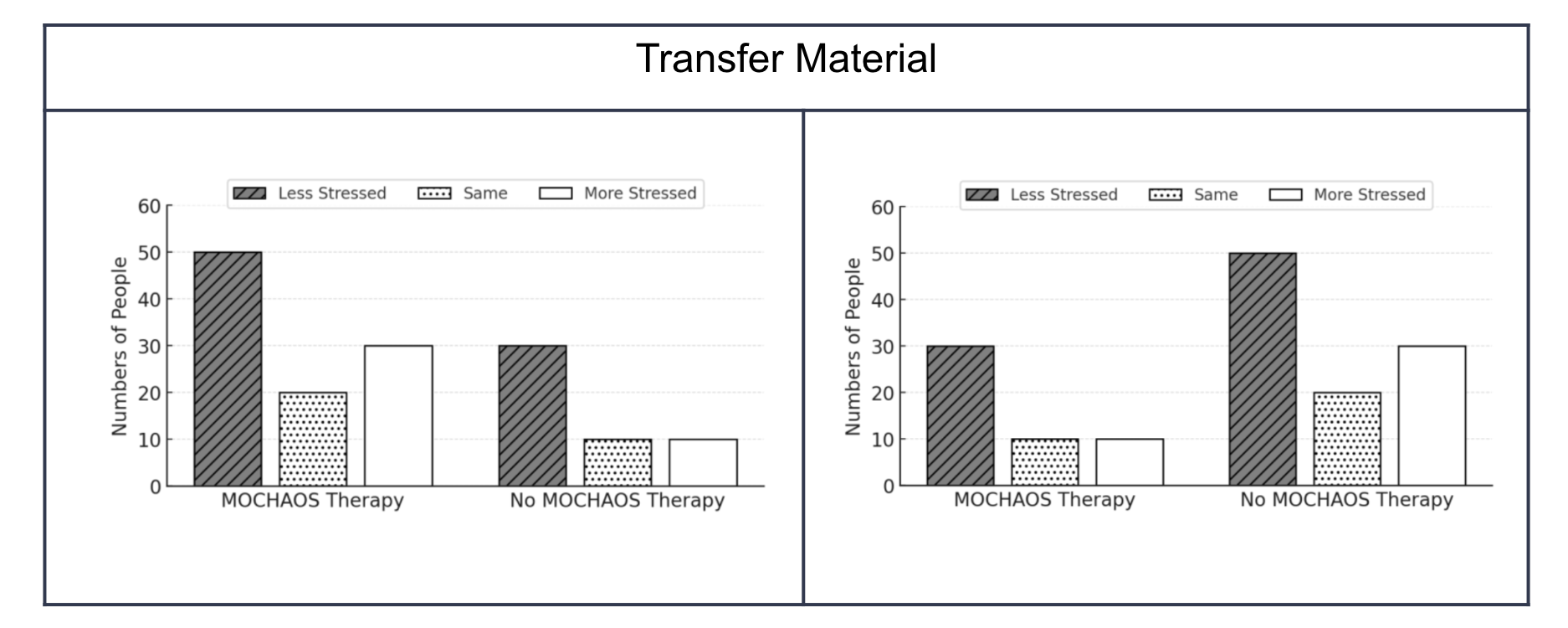}
  \caption{%
   Counterbalancing design for the transfer task%
  }
  \label{fig:transfer}
\end{figure}

The same numerical values were used in the PB and No-PB contexts, but the mapping between group labels and outcome values was counterbalanced across versions. For example, the right version of \cref{fig:transfer} reverses the group sizes and outcomes shown in the left version: 50 people received MOCHAOS therapy (30 \textit{Less Stressed}, 10 \textit{Same}, and 10 \textit{More Stressed}), whereas 100 people did not receive the therapy (50 \textit{Less Stressed}, 20 \textit{Same}, and 30 \textit{More Stressed}).

For each transfer question, participants were randomly assigned to read one of two counterbalanced versions, such that half received one configuration (e.g., higher likelihood of being more stressed for MOCHAOS Therapy; left of \cref{fig:transfer}) and half received the alternative configuration (e.g., lower likelihood of being more stressed for MOCHAOS Therapy; right of \cref{fig:transfer}). 
Participants were asked to infer the direction of the effect from the more complex bar chart, similar to the intervention task. 

Finally, participants completed a post-survey assessing their personal attitudes toward the PB topic using a Likert scale ranging from 0 (not helpful at all) to 10 (very helpful).

\rev{The transfer tasks were included to examine whether any benefit of the initial hand-drawing intervention would generalize. Unlike the intervention tasks, the transfer tasks did not include videos. The complexity of the visualizations also differed between the intervention task and the transfer task. 
The $2 \times 2$ bar charts in the intervention task were simpler to reason with than the $3 \times 3$ bar charts in the transfer task. 
We used bar charts throughout our study because they are one of the most widely used visualization types for comparison tasks, where misinterpretation and heuristic reasoning frequently occur~\cite{cleveland1984graphical, tversky1974judgment}. }


\subsection{Measures}

\noindent
\textbf{Response Accuracy}
Visualization comprehension was measured using participants’ responses to comparison questions in each task. 
For example, in the No-PB context, participants indicated whether patients who used the cream were more or less likely to have their rashes improve compared to those who did not. 
Responses were coded as a binary \textit{response accuracy} (0 = incorrect, 1 = correct) based on consistency with the depicted data.

\vspace{1mm}
\noindent
\textbf{Post Rating}
In the post-survey, participants reported their attitudes toward topics of PB context in each task, using an 11-point Likert scale ranging from 0 (not helpful at all) to 10 (very helpful). For example, regarding PB context of the main task (the handgun ban topic), participants were asked, ``In your personal opinion, to what extent do you think banning handguns is helpful for protecting personal safety?''

\vspace{1mm}
\noindent
\textbf{Response Type}
To examine how prior beliefs influenced responses, we constructed an ordinal variable, \textit{response type}, capturing the relationship between prior beliefs, comparison responses, and \textit{response accuracy}. Prior belief direction was derived from post ratings, with values greater than 5 indicating a \textit{more likely / effective} belief, less than 5 indicating a \textit{less likely / not effective} belief, and equal to 5 indicating no prior belief. 
Based on whether participants’ prior beliefs aligned with their selected response and whether the response was correct, we classified reasoning into four categories:
\begin{enumerate}[noitemsep, topsep=0pt]
    \item \textit{Occur}: belief-consistent and incorrect
    \item \textit{Contradict}: belief-inconsistent and incorrect
    \item \textit{Facilitate}: belief-consistent and correct
    \item \textit{Overcome}: belief-inconsistent and correct
\end{enumerate}
\rev{For example, if a participant believed that banning handguns decreases crime but correctly selected that banning handguns is \textbf{less} likely to decrease crime according to the chart, the response was classified as \textit{Overcome} because it was belief-inconsistent but correct.} These categories reflect increasing reasoning quality under the influence of prior beliefs, from belief-driven errors (\textit{Occur}) to correctly overriding prior beliefs (\textit{Overcome}). \rev{Participants reporting a neutral prior belief (rating = 5) were coded as belief-inconsistent because they did not express a directional expectation. We also repeated the analysis excluding these observations, and the pattern of results was unchanged.}

\begin{figure*}[t]
  \centering

  \begin{minipage}{0.45\textwidth}
    \centering
    \includegraphics[width=\linewidth]{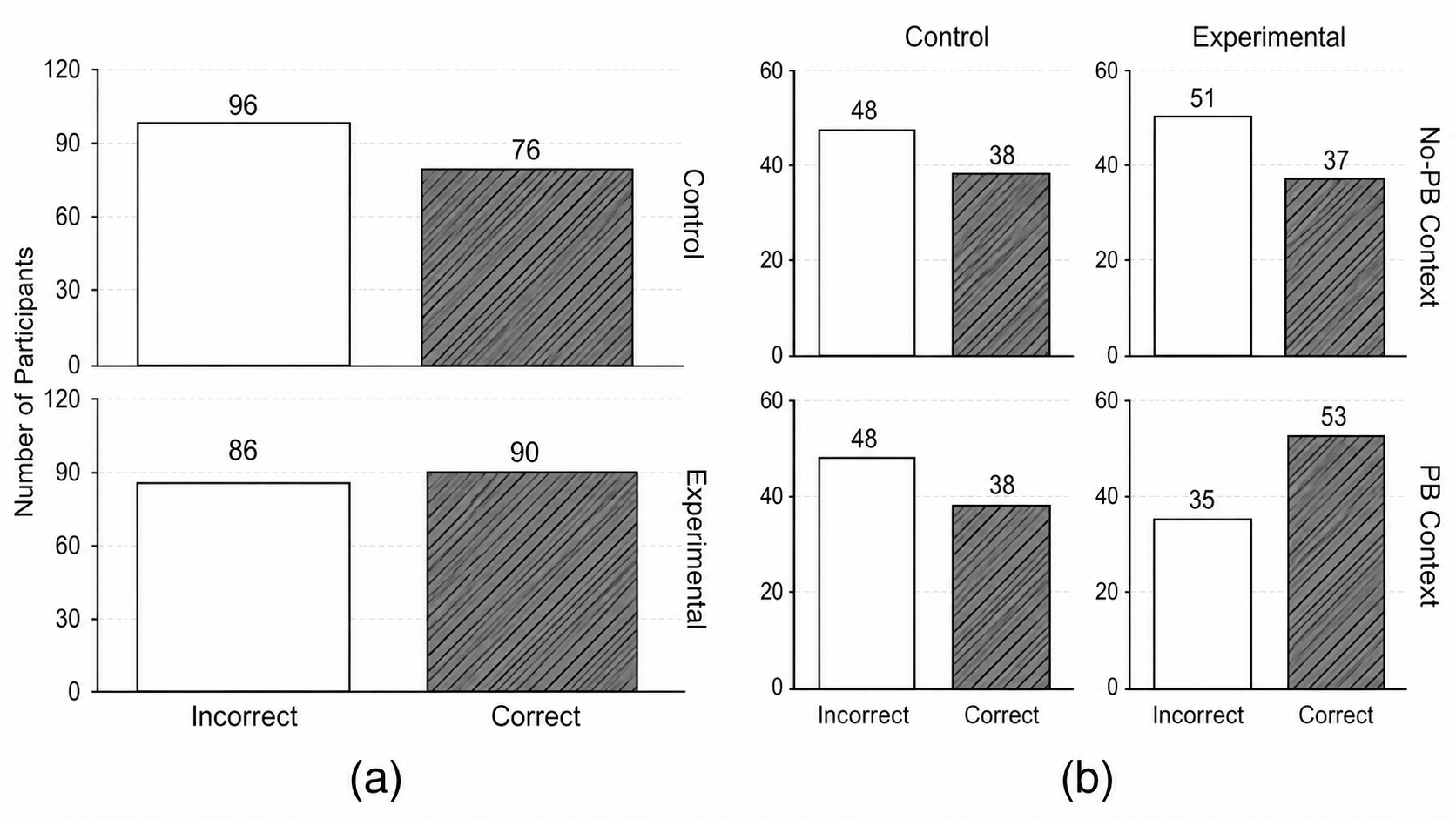}
  \end{minipage}
  \hfill
  \begin{minipage}{0.45\textwidth}
    \centering
    \includegraphics[width=\linewidth]{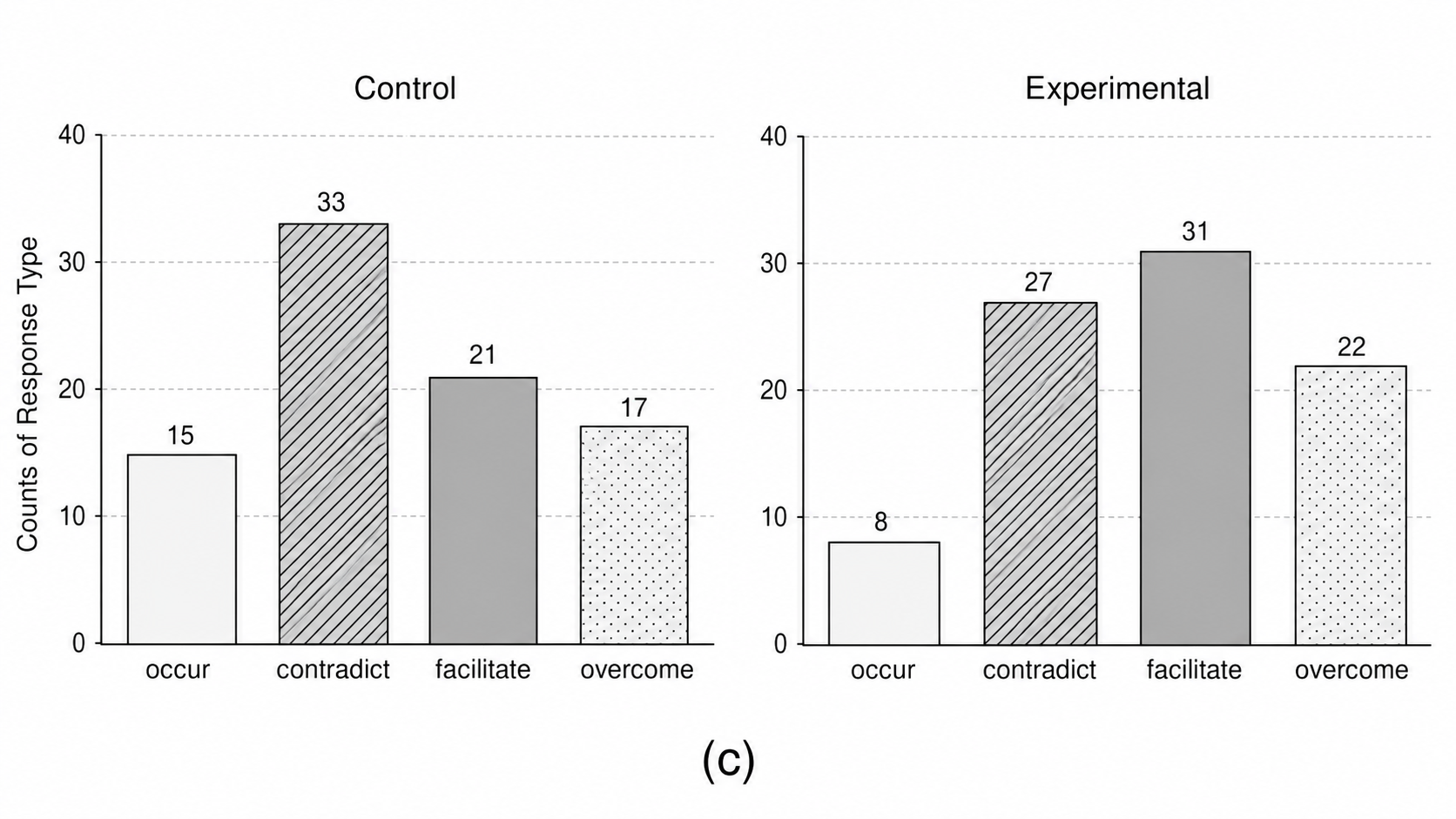}
  \end{minipage}

  \caption{
  (a) The distribution of Response Accuracy by condition;
  (b) The distribution of Response Accuracy by context;
  (c) The distribution of Response Type
  }
  \label{fig:all_fig}
\end{figure*}

\subsection{Results}
\cref{fig:all_fig} (a) shows the overall distribution of chart comprehension accuracy across two conditions. Participants in the control condition produced 76 correct responses (44.2\%) and 96 incorrect responses, whereas participants in the experimental condition produced 90 correct responses (51.1\%) and 86 incorrect responses. \rev{Unless otherwise noted, analyses included responses from both the intervention tasks and transfer tasks.}

\vspace{1mm}
\noindent
\textbf{H1: Effect of Embodied Construction.}
To evaluate whether embodied pedagogy improves interpretation accuracy, we fitted a generalized linear mixed-effects logistic regression with comprehension score as the outcome variable, condition (control or embodied) as a fixed effect, and random intercepts for participants and contexts (with prior belief or not). The main effect of condition was not statistically significant ($\beta = 0.29$, $SE = 0.23$, $z = 1.27$, $p = .204$), with an odds ratio of 1.34.

\vspace{1mm}
\noindent
\textbf{H2: Moderation by Prior-Belief.}
We next examined whether the effect of condition (control or embodied) differed across contexts (with prior belief or not). 
In the No-PB context (upper panel of \cref{fig:all_fig}(b)), the estimated coefficient for condition was $\beta = -0.09$ ($SE = 0.31$, $z = -0.29$, $p = .772$), with an odds ratio of 0.91. This suggests no improvement in accuracy under the embodied condition in the No-PB context.

In contrast, in the PB context (lower panel of \cref{fig:all_fig}(b)), the estimated coefficient for condition was $\beta = 0.70$ ($SE = 0.34$, $z = 2.03$, $p = .042$), with an odds ratio of 2.01.
This indicates that participants in the embodied condition had approximately \textbf{twice} the odds of producing a correct response compared to those in the control condition. This effect is statistically significant.

To formally test this difference, we fitted a model including the $condition \times context$ interaction. The interaction term was marginal ($\beta = 0.77$, $SE = 0.45$, $z = 1.73$, $p = .084$). See \cref{fig:all_fig}(b).

\vspace{1mm}
\noindent
\textbf{H3: Effects on Reasoning and Response Types.}
To examine how the intervention influenced reasoning, we analyzed the ordinal outcome \textit{response type} using a cumulative link mixed model with a random intercept for participants. The model revealed a significant main effect of condition ($\beta = 0.568$, $SE = 0.284$, $z = 2.00$, $p = .045$), with an odds ratio of 1.76.
This indicates that participants in the embodied condition had higher odds of producing \textit{Facilitate} and \textit{Overcome} types of responses than \textit{Occur} and \textit{Contradict}, which suggests more effective reasoning and data sense-making. 

Specifically, in the control condition, responses were most likely to be classified as \textit{Contradict} (38.81\%), followed by \textit{Facilitate} (27.82\%), \textit{Overcome} (17.16\%), and \textit{Occur} (16.22\%). In contrast, in the embodied condition, responses were most likely to fall into \textit{Facilitate} (32.29\%) or \textit{Contradict} (31.06\%), with a higher probability of \textit{Overcome} (26.76\%) and a lower probability of \textit{Occur} (9.89\%). See \cref{tab:cb_probabilities}.

\begin{table}[htbp]
\centering
\caption{Model-predicted probabilities of each response type by condition in the PB context}
\label{tab:cb_probabilities}

\begin{tabular}{lcc}
\toprule
Response Type & Control (\%) & Experimental (\%) \\
\midrule
\textit{Occur} & 16.22 & 9.89 \\
\textit{Contradict} & 38.81 & 31.06 \\
\textit{Facilitate} & 27.82 & 32.29 \\
\textit{Overcome} & 17.16 & 26.76 \\

\bottomrule
\end{tabular}

\end{table}

Pairwise comparisons further clarified these differences (\cref{fig:all_fig}(c)). The embodied condition showed a significantly higher probability of \textit{overcome} responses ($estimate = -0.096$, $SE = 0.048$, $z = -2.01$, $p = .045$), indicating that participants in the embodied condition were more likely to correctly interpret the visualization even when their prior beliefs suggested the opposite conclusion.

\section{Discussion}
We investigated whether watching dynamic hand-drawing of faceted bar charts can improve data interpretation, particularly when people have strong prior beliefs. 
Our results provide partial support for our hypotheses, consistent with the existing theoretical frameworks.
Although the embodied condition did not significantly improve accuracy across all contexts, it showed a clear advantage in belief-relevant contexts. When prior beliefs were more likely to bias interpretation, participants in the embodied condition were more likely to make the correct interpretation. 
\rev{The relatively high number of incorrect responses across conditions also suggests that the tasks were challenging for many participants. One likely reason is that the comparison questions required participants to reason about relative likelihoods rather than simply compare raw bar heights. In both the $2 \times 2$ intervention tasks and the more complex $3 \times 3$ transfer tasks, a correct response required integrating information across groups and outcomes. Participants may have relied on salient visual features, such as the tallest bar or the largest raw count, rather than comparing the proportional relationships represented in the chart. }

\rev{One possible explanation for the observed advantage in belief-relevant contexts is that the embodied, dynamic presentation changed how participants processed the visualization. Rather than immediately focusing on the most visually salient features, viewers may have been encouraged to attend to the relational structure of the graph, such as comparisons and proportional differences. In addition, the progressive, embodied unfolding of the visualization may have slowed processing, increasing the likelihood of more deliberative reasoning. This interpretation is consistent with the increase in \textit{Overcome} responses, suggesting that participants were more likely to override belief-consistent but incorrect intuitions.}

\rev{Although our study points to the potential of embodied approaches as a process-level intervention, our study does not isolate the effects of observing a hand draw the graph from those of the graph being progressively revealed over time. Our findings suggest that \textbf{dynamic graph construction}, which combines hand drawing with a gradual reveal of the visualization, can reduce belief-biased interpretation. Future work should disentangle these mechanisms by comparing hand-drawn and non-hand-drawn animated presentations.
}

\section{Limitations}

This study has several limitations. First, \rev{as we have discussed,} the design does not fully isolate the embodied component of the intervention. Some observed effects may be attributable to the dynamic presentation of the visualization rather than embodiment itself. \rev{The video stimuli also involved several production choices that may have influenced participants' interpretation. For example, the hand-drawing videos used a printed template to keep the final chart visually consistent with the static control chart. Although this helped control the final visual appearance across conditions, it also meant that some chart structure was visible before being traced, which may have reduced the distinctiveness of the hand-drawing manipulation.}

Second, the study was limited to bar charts in a fixed order. Future work can adopt counterbalanced designs to control for order effects and expand the range of tasks. 

\rev{Third}, the intervention was evaluated on relatively simple visualizations ($2 \times 2$ and $3 \times 3$ bar charts). Extending this approach to more complex visualizations, such as jitter plots, will be important for understanding whether the observed benefits persist across different forms of reasoning.

\rev{Finally, our coding scheme grouped neutral responses into the belief-inconsistent category rather than treating them as a distinct group. As a result, the \textit{Contradict} and \textit{Overcome} categories should not be strictly interpreted as overriding a directional prior belief. Future work can isolate or exclude neutral responses to refine these analyses.}



\bibliographystyle{abbrv-doi-hyperref}

\bibliography{ConfimationBiasRefer}

\end{document}